\documentclass[12pt,titlepage,letterpaper]{utarticle}

\usepackage{amsmath}
\usepackage{amsfonts}
\usepackage{amssymb}
\usepackage{amsthm}
\usepackage{xparse}
\usepackage{mathtools}
\usepackage{graphicx}
\usepackage{color}
\usepackage[utf8x]{inputenc}
\usepackage{tocloft}
\usepackage[titletoc,title]{appendix}
\usepackage{cite}
\usepackage{hyperref}
\usepackage{longtable}

\definecolor{aqua}{rgb}{0, 1.0, 1.0}
\definecolor{fuschia}{rgb}{1.0, 0, 1.0}
\definecolor{gray}{rgb}{0.502, 0.502, 0.502}
\definecolor{lime}{rgb}{0, 1.0, 0}
\definecolor{maroon}{rgb}{0.502, 0, 0}
\definecolor{navy}{rgb}{0, 0, 0.502}
\definecolor{olive}{rgb}{0.502, 0.502, 0}
\definecolor{purple}{rgb}{0.502, 0, 0.502}
\definecolor{silver}{rgb}{0.753, 0.753, 0.753}
\definecolor{teal}{rgb}{0, 0.502, 0.502}

\makeatletter
\newdimen\itex@wd%
\newdimen\itex@dp%
\newdimen\itex@thd%
\def\itexspace#1#2#3{\itex@wd=#3em%
\itex@wd=0.1\itex@wd%
\itex@dp=#2ex%
\itex@dp=0.1\itex@dp%
\itex@thd=#1ex%
\itex@thd=0.1\itex@thd%
\advance\itex@thd\the\itex@dp%
\makebox[\the\itex@wd]{\rule[-\the\itex@dp]{0cm}{\the\itex@thd}}}
\makeatother

\makeatletter
\newif\if@sup
\newtoks\@sups
\def\append@sup#1{\edef\act{\noexpand\@sups={\the\@sups #1}}\act}%
\def\reset@sup{\@supfalse\@sups={}}%
\def\mk@scripts#1#2{\if #2/ \if@sup ^{\the\@sups}\fi \else%
  \ifx #1_ \if@sup ^{\the\@sups}\reset@sup \fi {}_{#2}%
  \else \append@sup#2 \@suptrue \fi%
  \expandafter\mk@scripts\fi}
\def\tensor#1#2{\reset@sup#1\mk@scripts#2_/}
\def\multiscripts#1#2#3{\reset@sup{}\mk@scripts#1_/#2%
  \reset@sup\mk@scripts#3_/}
\makeatother

\makeatletter
\newbox\slashbox \setbox\slashbox=\hbox{$/$}
\def\itex@pslash#1{\setbox\@tempboxa=\hbox{$#1$}
  \@tempdima=0.5\wd\slashbox \advance\@tempdima 0.5\wd\@tempboxa
  \copy\slashbox \kern-\@tempdima \box\@tempboxa}
\def\slash{\protect\itex@pslash}
\makeatother

\def\clap#1{\hbox to 0pt{\hss#1\hss}}

\let\oldroot\root
\def\root#1#2{\oldroot #1 \of{#2}}
\renewcommand{\sqrt}[2][]{\oldroot #1 \of{#2}}

\DeclareSymbolFont{symbolsC}{U}{txsyc}{m}{n}
\SetSymbolFont{symbolsC}{bold}{U}{txsyc}{bx}{n}
\DeclareFontSubstitution{U}{txsyc}{m}{n}

\DeclareSymbolFont{stmry}{U}{stmry}{m}{n}
\SetSymbolFont{stmry}{bold}{U}{stmry}{b}{n}

\DeclareFontFamily{OMX}{MnSymbolE}{}
\DeclareSymbolFont{mnomx}{OMX}{MnSymbolE}{m}{n}
\SetSymbolFont{mnomx}{bold}{OMX}{MnSymbolE}{b}{n}
\DeclareFontShape{OMX}{MnSymbolE}{m}{n}{
    <-6>  MnSymbolE5
   <6-7>  MnSymbolE6
   <7-8>  MnSymbolE7
   <8-9>  MnSymbolE8
   <9-10> MnSymbolE9
  <10-12> MnSymbolE10
  <12->   MnSymbolE12}{}

\makeatletter
\def\re@DeclareMathSymbol#1#2#3#4{%
    \let#1=\undefined
    \DeclareMathSymbol{#1}{#2}{#3}{#4}}
\re@DeclareMathSymbol{\neArrow}{\mathrel}{symbolsC}{116}
\re@DeclareMathSymbol{\neArr}{\mathrel}{symbolsC}{116}
\re@DeclareMathSymbol{\seArrow}{\mathrel}{symbolsC}{117}
\re@DeclareMathSymbol{\seArr}{\mathrel}{symbolsC}{117}
\re@DeclareMathSymbol{\nwArrow}{\mathrel}{symbolsC}{118}
\re@DeclareMathSymbol{\nwArr}{\mathrel}{symbolsC}{118}
\re@DeclareMathSymbol{\swArrow}{\mathrel}{symbolsC}{119}
\re@DeclareMathSymbol{\swArr}{\mathrel}{symbolsC}{119}
\re@DeclareMathSymbol{\nequiv}{\mathrel}{symbolsC}{46}
\re@DeclareMathSymbol{\Perp}{\mathrel}{symbolsC}{121}
\re@DeclareMathSymbol{\Vbar}{\mathrel}{symbolsC}{121}
\re@DeclareMathSymbol{\sslash}{\mathrel}{stmry}{12}
\re@DeclareMathSymbol{\bigsqcap}{\mathop}{stmry}{"64}
\re@DeclareMathSymbol{\biginterleave}{\mathop}{stmry}{"6}
\re@DeclareMathSymbol{\invamp}{\mathrel}{symbolsC}{77}
\re@DeclareMathSymbol{\parr}{\mathrel}{symbolsC}{77}
\makeatother

\makeatletter
\def\Decl@Mn@Delim#1#2#3#4{%
  \if\relax\noexpand#1%
    \let#1\undefined
  \fi
  \DeclareMathDelimiter{#1}{#2}{#3}{#4}{#3}{#4}}
\def\Decl@Mn@Open#1#2#3{\Decl@Mn@Delim{#1}{\mathopen}{#2}{#3}}
\def\Decl@Mn@Close#1#2#3{\Decl@Mn@Delim{#1}{\mathclose}{#2}{#3}}
\Decl@Mn@Open{\llangle}{mnomx}{'164}
\Decl@Mn@Close{\rrangle}{mnomx}{'171}
\Decl@Mn@Open{\lmoustache}{mnomx}{'245}
\Decl@Mn@Close{\rmoustache}{mnomx}{'244}
\makeatother

\makeatletter
\DeclareRobustCommand\widecheck[1]{{\mathpalette\@widecheck{#1}}}
\def\@widecheck#1#2{%
    \setbox\z@\hbox{\m@th$#1#2$}%
    \setbox\tw@\hbox{\m@th$#1%
       \widehat{%
          \vrule\@width\z@\@height\ht\z@
          \vrule\@height\z@\@width\wd\z@}$}%
    \dp\tw@-\ht\z@
    \@tempdima\ht\z@ \advance\@tempdima2\ht\tw@ \divide\@tempdima\thr@@
    \setbox\tw@\hbox{%
       \raise\@tempdima\hbox{\scalebox{1}[-1]{\lower\@tempdima\box
\tw@}}}%
    {\ooalign{\box\tw@ \cr \box\z@}}}
\makeatother

\makeatletter
\NewDocumentCommand\mathraisebox{moom}{%
\IfNoValueTF{#2}{\def\@temp##1##2{\raisebox{#1}{$\m@th##1##2$}}}{%
\IfNoValueTF{#3}{\def\@temp##1##2{\raisebox{#1}[#2]{$\m@th##1##2$}}%
}{\def\@temp##1##2{\raisebox{#1}[#2][#3]{$\m@th##1##2$}}}}%
\mathpalette\@temp{#4}}
\makeatletter

\makeatletter
\def\udots{\mathinner{\mkern2mu\raise\p@\hbox{.}
\mkern2mu\raise4\p@\hbox{.}\mkern1mu
\raise7\p@\vbox{\kern7\p@\hbox{.}}\mkern1mu}}
\makeatother

\theoremstyle{plain}

\theoremstyle{definition}

\theoremstyle{remark}

\begin{document}

\preprint{
UTTG--06--18\\
}

\title{Product SCFTs for the $E_7$ Theory}

\author{Jacques Distler and Behzat Ergun
     \oneaddress{
      Theory Group\\
      Department of Physics,\\
      University of Texas at Austin,\\
      Austin, TX 78712, USA \\
      {~}\\
      \email{distler@golem.ph.utexas.edu}\\
      \email{bergun@utexas.edu}\\
      }
}
\date{March 7, 2018}

\Abstract{We present a simple criterion for when an $\mathcal{N}=2$ SCFT must be a product SCFT. Applied to the class-S theories of type $E_7$, we find 29 (out of 11,000) 3-punctured spheres which are product SCFTs.
}

\maketitle

\tocloftpagestyle{empty}
\tableofcontents
\vfill
\newpage
\setcounter{page}{1}

\section{Introduction}\label{introduction}

In the past decade, the class-S construction \cite{Gaiotto:2009hg,Gaiotto:2009we} has furthered our understanding of 4d $\mathcal{N}=2$ supersymmetric theories and their superconformal fixed points. Generically, $\mathcal{N}=2$ SCFTs occur in families, parameterized by exactly-marginal deformations corresponding to complexified gauge couplings $\tau_i\sim \tfrac{\theta_i}{\pi}+\tfrac{8\pi i}{g_i^2}$. Upon turning off these gauge couplings, one is left with some free vector multiplets and an isolated SCFT whose global symmetry group contains the previously gauged group as a subgroup. Thus, to classify such theories, one must classify isolated SCFTs and their possible gaugings.

Class-S theories are obtained by partially-twisted compactification of a 6d $(2,0)$ theory on a genus-$g$, $n$-punctured Riemann surface $\mathcal{C}_{g,n}$. The exactly-marginal deformations correspond to deforming the complex structure of $\mathcal{C}_{g,n}$. Each pants-decomposition of the surface gives a presentation of this family of SCFTs as a gauge theory. Turning off all the gauge couplings corresponds to degenerating $\mathcal{C}_{g,n}$ into a nodal curve whose normalization consists of $(2g-2+n)$ 3-punctured spheres. The isolated SCFTs, corresponding to 3-punctured spheres (``fixtures''), and their gaugings are the basic ingredients in the classification of theories of class-S. Fixtures can be broadly categorized into three subsets: free hypermultiplets, isolated interacting SCFTs and a mixture of both.

For any given $(2,0)$ theory of a particular ADE type, there is a finite list of fixtures. In fact, these lists have redundancies, in the sense that many isolated SCFTs have multiple realizations as fixtures. Furthermore, some of the fixtures correspond to product theories, as noted in \cite{Chacaltana:2011ze,Chacaltana:2012ch,Chacaltana:2014jba,Chacaltana:2015bna,Chacaltana:2017boe} and investigated systematically in our previous work \cite{Distler:2017xba}.

Our previous results on identifying product theories involved computing unrefined Schur and Hall-Littlewood(HL) indices\cite{Rastelli:2014jja,Kinney:2005ej,Gadde:2011uv} for the fixtures. The main bottleneck in that computation lies in constructing the HL polynomials which scale with the size of the Weyl group and are considerably more difficult to compute in the case of $E_7$, $E_8$.

In this work, we present an alternative criterion for identifying product theories --- which stems from the unitarity bounds on the allowed values of levels and central charges \cite{Beem:2013sza,Lemos:2015orc}. Our criterion provides a sufficient (but possibly not necessary) condition for a class-S theory to be a product SCFT. Nevertheless, applying it, we recover \emph{all} of the fixtures which were previously-known to be product SCFTs, along with a few new additions to our list. In particular, of the 11,000 good fixtures in the $E_7$ theory, we find 29 product SCFTs. The same analysis was applied in \cite{Chcaltana:2018zag} to the $E_8$ theory, yielding 29 product SCFTs among the 49,836 fixtures of that theory.

In addition to being computationally simple, our criterion provides a conceptual understanding for \emph{why} a given fixture turns out to be a product SCFT: given the global symmetry and levels, the product structure is \emph{required} by unitarity. It also explains the ubiquity of Minahan-Nemeschansy (MN) theories \cite{Minahan:1996fg,Minahan:1996cj} as factors in the product SCFTs which arise in this way: the global symmetries and levels of the rank-1 MN theories lie on the boundary of the allowed region and so -- when they appear -- always decompose as a factor in the product. In turn, higher-rank MN theories, though they do not themselves saturate the unitarity bound, are often related to products of rank-1 MN theories by the Sommers-Achar group action on one of the punctures (see e.g.~\S3.1 of \cite{Chcaltana:2018zag}).

\section{Criterion for Decomposability from Unitarity}\label{criterion_for_decomposability_from_unitarity}

Let $F=(F_1)_{k_1}\times \cdots \times(F_n)_{k_n}$ be the flavor symmetry for a theory. For each simple factor, $F_i$, in the global symmetry group, there is a lower bound on the level, $k_i$, which follows from unitarity. If the theory is indecomposable (possessing a single stress tensor), then \cite{Beem:2013sza}

\begin{equation}
k_i \geq \frac{24 \kappa_{F_i}c}{\dim(F_i)+12 c}
\label{unitarityklowerbound}\end{equation}
where $\kappa_F$ is the dual Coxeter number and $c=(2n_v+n_h)/12$ is the Weyl-anomaly coefficient. In examining the global symmetries of a fixture, if this bound should happen to be violated for one of the $F_i$, then it must be that the assumption of only a single $\hat{C}_{0(0,0)}$ (stress-tensor) multiplet must be wrong. Hence, the theory \emph{must} be a product SCFT, with $c=\sum c_i$, and the global symmetries distributed between the factors so that the bound is satisfied within each factor.

\noindent
Using this criterion, we can immediately arrive at several important conclusions.

First of all, in searching for product SCFTs, we can restrict our attention to fixtures which have an enhanced global symmetry. For the global symmetry associated to any puncture $P$, \eqref{unitarityklowerbound} is satisfied for the fixture consisting of $P$ and two full punctures. Since the RHS of \eqref{unitarityklowerbound} is a monotonically-increasing function of $c$, replacing the full punctures by any other punctures (which decreases $c$ for the fixture) leaves the bound still satisfied.

Second, we note that the rank-1 Minahan-Nemeschansky theories with flavor symmetries ${(E_6)}_6$, ${(E_7)}_8$ and ${(E_8)}_{12}$ saturate the bound \eqref{unitarityklowerbound}. They also saturate the bound \cite{Lemos:2015orc}

\begin{equation}
k_i(-180c^2+ 66c + 3\dim(F_i))+(60c^2-22c)\kappa_{F_i}\leq 0
\label{unitarityupperbound}\end{equation}
Together, \eqref{unitarityklowerbound} and \eqref{unitarityupperbound} imply that the value of $k$ for the rank-1 Minahan-Nemeschansky theory is the \emph{lowest one} compatible with unitarity and that, for this value of $k$, there is a unique value of $c$ compatible with unitarity. More generally, the allowed values of $c$ and $k$ are the shaded regions in the graph below (respectively for $E_6$, $E_7$ and $E_8$), with the rank-1 Minahan-Nemeschansky at the ``corner'' of the allowed region (respectively at $(k,c)=\left(6, 13/6\right)$,  $\left(8, 19/6\right)$ and $\left(12,31/6\right)$).

\begin{displaymath}
 \includegraphics[width=183pt]{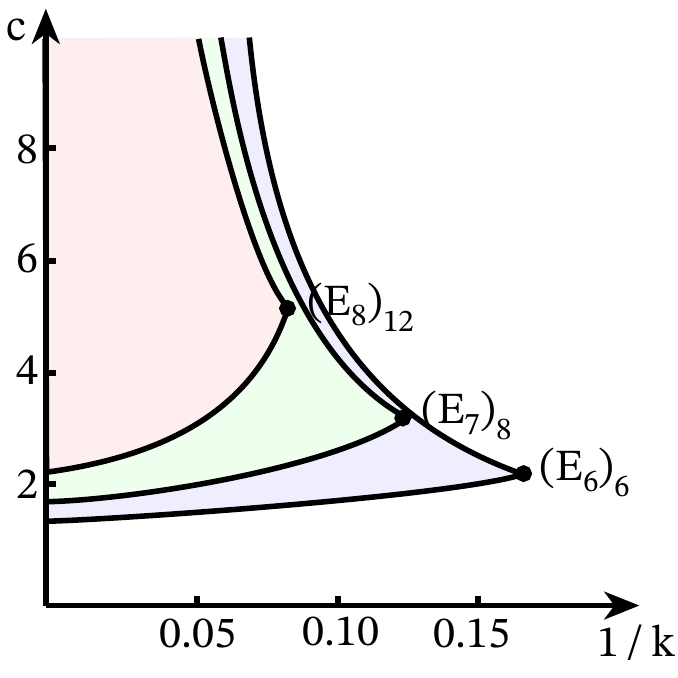}
\end{displaymath}
The upshot is that \emph{any} appearance of ${(E_6)}_6$, ${(E_7)}_8$ or ${(E_8)}_{12}$ in the global symmetry \emph{must} correspond to a decoupled rank-1 Minahan-Nemeschansky.

While a violation of \eqref{unitarityklowerbound} is a \emph{sufficient} condition to prove that a theory is product SCFT, we have no proof that it is \emph{necessary}. Nevertheless, all of the \emph{known} product SCFTs, both in our previous work \cite{Distler:2017xba} and in \cite{Chacaltana:2017boe} can be deduced from this simple criterion.

Let us turn, then, to applying this criterion to the $E_7$ theory \cite{Chacaltana:2017boe}.

\section{Enhanced ${(E_7)}_{36}$}\label{enhanced_}

A simple class of examples arise when the ${(E_7)}_{36}$ symmetry of the full puncture is enhanced to ${(E_7)}_{36-k}\times {(E_7)}_{k}$. In this case, the level of one of the two $(E_7)$s must be $\leq 18$. But $k=18$ violates the inequality \eqref{unitarityklowerbound} for the value of $c$ of the fixture. Hence the theory must be a product.

{
\footnotesize
\renewcommand{\arraystretch}{2.25}

\begin{longtable}{|c|c|c|c|c|}
\hline
\#&Fixture&Graded Coulomb Branch Dims&$(n_h,n_v)$&Global Symmetry\\
\hline 
\endhead
1&$\begin{matrix} \includegraphics[width=76pt]{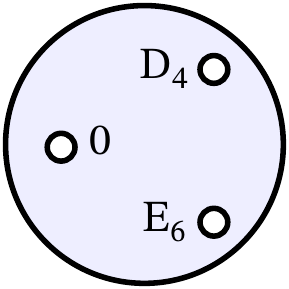}\end{matrix}$&$\{0, 0, 2, 0, 2, 1, 0, 2, 0, 1, 0\}$&$(204, 116)$&${(E_7)}_{36-k}\times {(E_7)}_{k}\times {Sp(3)}_{12}\times {SU(2)}_{12}$\\
\hline 
2&$\begin{matrix} \includegraphics[width=76pt]{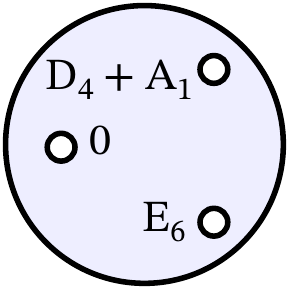}\end{matrix}$&$\{0, 0, 2, 0, 1, 1, 0, 2, 0, 1, 0\}$&$(190, 105)$&${(E_7)}_{36-k}\times {(E_7)}_{k}\times {Sp(2)}_{11}\times {SU(2)}_{12}$\\
\hline 
3&$\begin{matrix} \includegraphics[width=76pt]{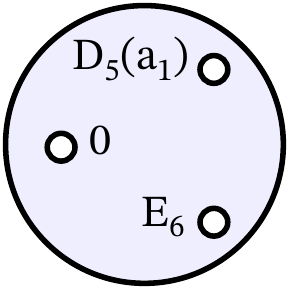}\end{matrix}$&$\{0, 0, 2, 1, 1, 1, 0, 1, 0, 1, 0\}$&$(178, 95)$&$\begin{gathered} {(E_7)}_{36-k}\times {(E_7)}_{k}\\ \times\\ {SU(2)}_{10}\times U(1)\times {SU(2)}_{12}\end{gathered}$\\
\hline 
4&$\begin{matrix} \includegraphics[width=76pt]{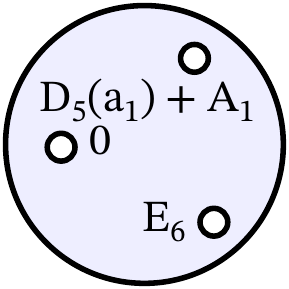}\end{matrix}$&$\{0, 0, 2, 0, 1, 1, 0, 1, 0, 1, 0\}$&$(168, 86)$&${(E_7)}_{36-k}\times {(E_7)}_{k}\times {SU(2)}_{56}\times {SU(2)}_{12}$\\
\hline 
5&$\begin{matrix} \includegraphics[width=76pt]{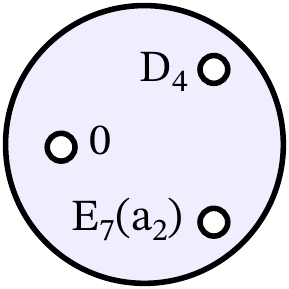}\end{matrix}$&$\{0, 0, 2, 0, 1, 1, 0, 2, 0, 1, 0\}$&$(192, 105)$&${(E_7)}_{36-k}\times {(E_7)}_{k}\times {Sp(3)}_{12}$\\
\hline 
6&$\begin{matrix} \includegraphics[width=76pt]{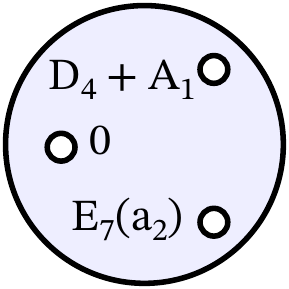}\end{matrix}$&$\{0, 0, 2, 0, 0, 1, 0, 2, 0, 1, 0\}$&$(178, 94)$&${(E_7)}_{36-k}\times {(E_7)}_{k}\times {Sp(2)}_{11}$\\
\hline 
7&$\begin{matrix} \includegraphics[width=76pt]{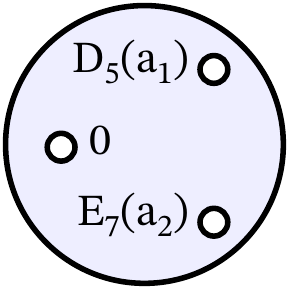}\end{matrix}$&$\{0, 0, 2, 1, 0, 1, 0, 1, 0, 1, 0\}$&$(166, 84)$&${(E_7)}_{36-k}\times {(E_7)}_{k}\times {SU(2)}_{10}\times U(1)$\\
\hline 8&$\begin{matrix} \includegraphics[width=76pt]{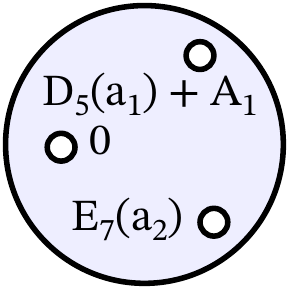}\end{matrix}$&$\{0, 0, 2, 0, 0, 1, 0, 1, 0, 1, 0\}$&$(156, 75)$&${(E_7)}_{36-k}\times {(E_7)}_{k}\times {SU(2)}_{56}$\\
\hline
\end{longtable}
}

\noindent
where the graded Coulomb Branch dimensions listed are $\{n_{2},n_{3},n_{4},n_{5},n_{6},n_{8},n_{9},n_{10},n_{12},n_{14},n_{18}\}$.

\#3 is the product of the ${(E_7)}_{8}$ Minahan-Nemeschansky theory with a rank-6 theory with global symmetry ${(E_7)}_{28}\times {SU(2)}_{12}\times{SU(2)}_{10}\times U(1)$ and $(n_h,n_v)=(154, 88)$. \#7 is the product of the ${(E_7)}_{8}$ Minahan-Nemeschansky theory with a rank-5 theory with global symmetry ${(E_7)}_{28}\times{SU(2)}_{10}\times U(1)$ and $(n_h,n_v)=(142, 77)$. Both of these new theories have realizations in the $A_{13}$ theory as the fixtures

\begin{displaymath}
 \includegraphics[width=273pt]{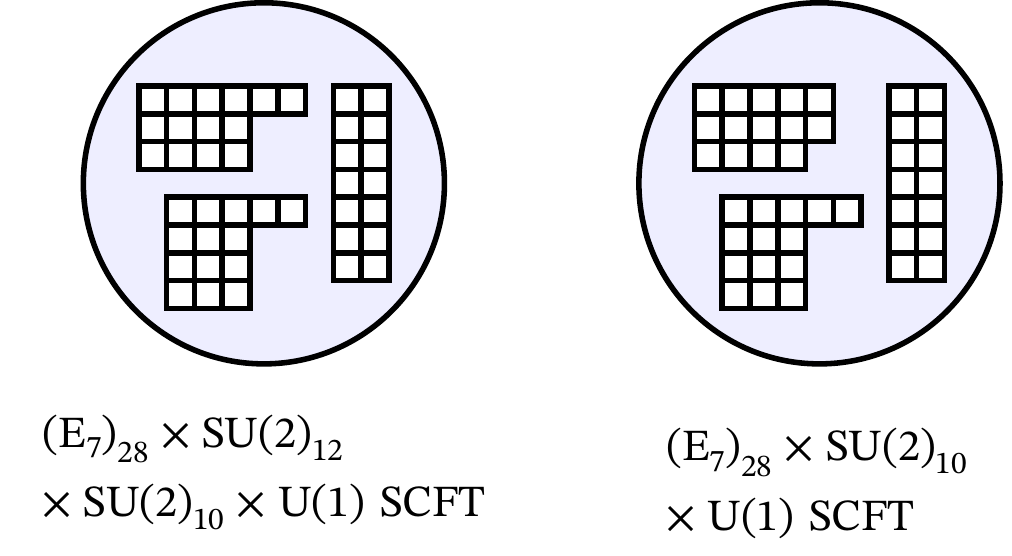}
\end{displaymath}
The other entries in the table are all consistent with one of the factors in the product being the ${(E_7)}_{8}$ Minahan-Nemeschansky theory, but we don't have alternative realizations of the other factor in the product.

\section{Known Product Theories}\label{known_product_theories}

The same consideration applies to other fixtures with enhanced global symmetries. There are 21 more fixtures in the $E_7$ theory where the unitarity bound forces the theory to be a product SCFT. Happily, all of these either contain rank-1 Minahan-Nemeschansky factors or decompose into products of known SCFTs.

{
\footnotesize
\renewcommand{\arraystretch}{2.25}

\begin{longtable}{|c|c|c|c|c|}
\hline
\#&Fixture&\mbox{\shortstack{Graded Coulomb\\ Branch Dims}}&$(n_h,n_v)$&Theory\\
\hline 
\endhead
1&$\begin{matrix} \includegraphics[width=76pt]{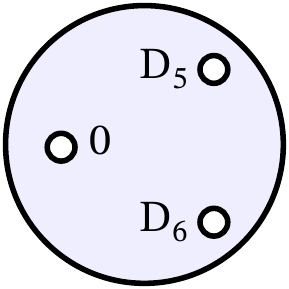}\end{matrix}$&$\{0, 0, 1, 0, 3, 1, 0, 0, 1, 0, 0\}$&$(159, 78)$&$\begin{gathered} [{(E_8)}_{12}\,\text{SCFT}]\\ \times\\ [{(E_7)}_{24}\times {SU(2)}_{8}\times {SU(2)}_{7}\,\text{SCFT}]\end{gathered}$\\
\hline 
2&$\begin{matrix} \includegraphics[width=76pt]{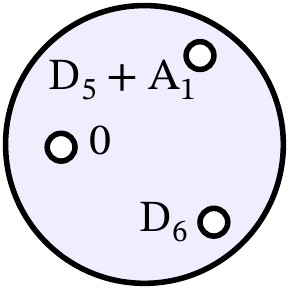}\end{matrix}$&$\{0, 0, 0, 0, 3, 1, 0, 0, 1, 0, 0\}$&$(151, 71)$&$[{(E_8)}_{12}\,\text{SCFT}]\times [{(E_7)}_{24}\times {SU(2)}_{7}\,\text{SCFT}]$\\
\hline 
3&$\begin{matrix} \includegraphics[width=76pt]{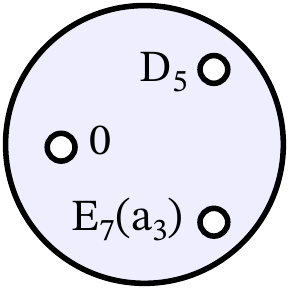}\end{matrix}$&$\{0, 1, 1, 0, 2, 1, 0, 0, 1, 0, 0\}$&$(152, 72)$&$[{(E_8)}_{12}\,\text{SCFT}]\times[{(E_7)}_{24}\times {SU(2)}_{8}\,\text{SCFT}]$\\
\hline 
4&$\begin{matrix} \includegraphics[width=76pt]{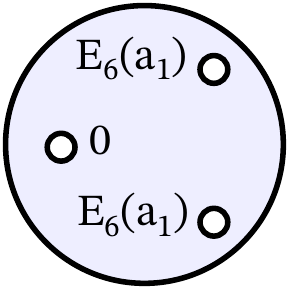}\end{matrix}$&$\{0, 2, 0, 2, 0, 0, 2, 0, 0, 0, 0\}$&$(140, 62)$&${[{(E_7)}_{18}\times U(1)\,\text{SCFT}]}^2$\\
\hline
5&$\begin{matrix} \includegraphics[width=76pt]{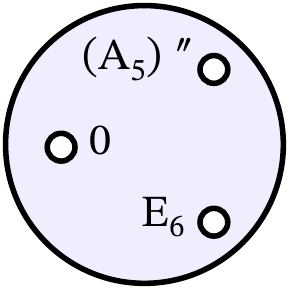}\end{matrix}$&$\{0, 0, 1, 0, 2, 1, 0, 0, 1, 0, 0\}$&$(152, 67)$&$[{(E_8)}_{12}\,\text{SCFT}]\times [{(E_7)}_{24}\times {(G_2)}_{12}\,\text{SCFT}]$\\
\hline 
6&$\begin{matrix} \includegraphics[width=76pt]{E7th0D5a1E6}\end{matrix}$&$\{0, 0, 2, 1, 1, 1, 0, 1, 0, 1, 0\}$&$(178, 95)$&$\begin{gathered} [{(E_7)}_{8}\,\text{SCFT}]\\ \times\\ [{(E_7)}_{28}\times {SU(2)}_{12}\times {SU(2)}_{10}\times U(1)\,\text{SCFT}]\end{gathered}$\\
\hline
7&$\begin{matrix} \includegraphics[width=76pt]{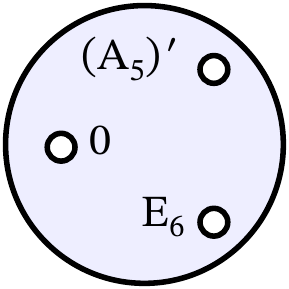}\end{matrix}$&$\{0, 0, 2, 0, 1, 1, 0, 1, 0, 0, 0\}$&$(141, 59)$&$\begin{gathered} [{(E_7)}_{16}\times {SU(2)}_{9}\,\text{SCFT}]\\ \times\\ [{(E_7)}_{20}\times {SU(2)}_{20}\times {SU(2)}_{12}\,  \text{SCFT}]\end{gathered}$\\
\hline
8&$\begin{matrix} \includegraphics[width=76pt]{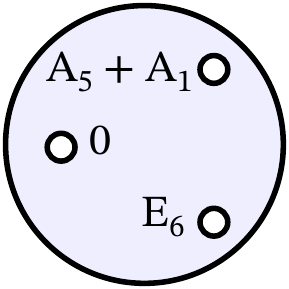}\end{matrix}$&$\{0, 0, 1, 0, 1, 1, 0, 0, 1, 0, 0\}$&$(138, 56)$&$[{(E_8)}_{12}\,\text{SCFT}]\times [{(E_7)}_{24}\times {SU(2)}_{26}\,\text{SCFT}]$\\
\hline
9&$\begin{matrix} \includegraphics[width=76pt]{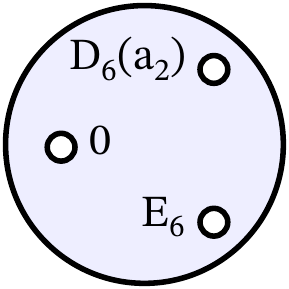}\end{matrix}$&$\{0, 0, 2, 0, 1, 1, 0, 0, 0, 0, 0\}$&$(121, 40)$&$\begin{gathered} [{(E_8)}_{12}\,\text{SCFT}]\\ \times\\  [{(E_7)}_{8}\,\text{SCFT}]\times [{(E_7)}_{16}\times {SU(2)}_{9}\,\text{SCFT}]\end{gathered}$\\
\hline
10&$\begin{matrix} \includegraphics[width=76pt]{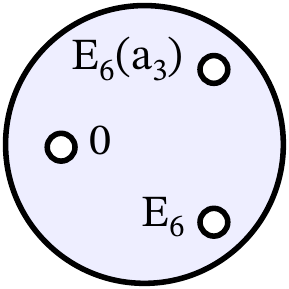}\end{matrix}$&$\{0, 0, 3, 0, 1, 0, 0, 1, 0, 0, 0\}$&$(132, 51)$&$\begin{gathered} {[{(E_7)}_{8}\,\text{SCFT}]}^2\\ \times\\ [{(E_7)}_{20}\times SU(2)_{20}\times SU(2)_{12}\,\text{SCFT}]\end{gathered}$\\
\hline 
11&$\begin{matrix} \includegraphics[width=76pt]{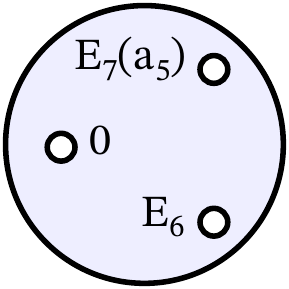}\end{matrix}$&$\{0, 0, 3, 0, 1, 0, 0, 0, 0, 0, 0\}$&$(112, 32)$&$[{(E_8)}_{12}\,\text{SCFT}]\times {[{(E_7)}_{8}\,\text{SCFT}]}^3$\\
\hline 
12&$\begin{matrix} \includegraphics[width=76pt]{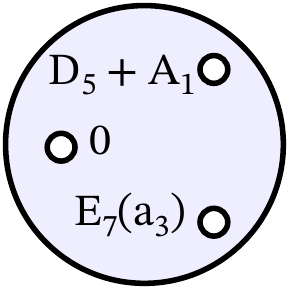}\end{matrix}$&$\{0, 1, 0, 0, 2, 1, 0, 0, 1, 0, 0\}$&$(144, 65)$&$[{(E_8)}_{12}\,\text{SCFT}]\times[{(E_7)}_{24}\,\text{SCFT}]$\\
\hline 
13&$\begin{matrix} \includegraphics[width=76pt]{E7th0D5a1E7a2}\end{matrix}$&$\{0, 0, 2, 1, 0, 1, 0, 1, 0, 1, 0\}$&$(166, 84)$&$\begin{gathered} [{(E_7)}_{8}\,\text{SCFT}]\\ \times\\ [{(E_7)}_{28}\times {SU(2)}_{10}\times U(1)\,\text{SCFT}]\end{gathered}$\\
\hline 
14&$\begin{matrix} \includegraphics[width=76pt]{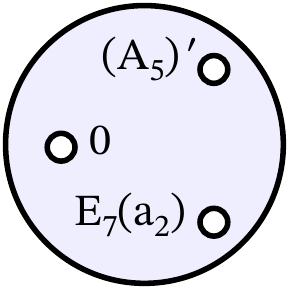}\end{matrix}$&$\{0, 0, 2, 0, 0, 1, 0, 1, 0, 0, 0\}$&$(129, 48)$&$[{(E_8)}_{20}\,\text{SCFT}]\times [{(E_7)}_{16}\times {SU(2)}_9\,\text{SCFT}]$\\
\hline 
15&$\begin{matrix} \includegraphics[width=76pt]{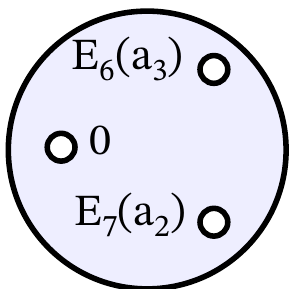}\end{matrix}$&$\{0, 0, 3, 0, 0, 0, 0, 1, 0, 0, 0\}$&$(120, 40)$&$[{(E_8)}_{20}\,\text{SCFT}]\times {[{(E_7)}_{8}\,\text{SCFT}]}^2$\\
\hline 
16&$\begin{matrix} \includegraphics[width=76pt]{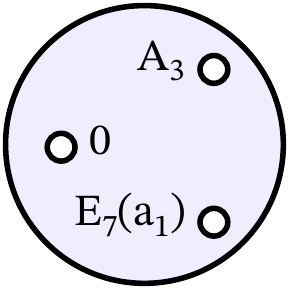}\end{matrix}$&$\{0, 0, 0, 0, 2, 1, 0, 0, 1, 0, 0\}$&$(152, 60)$&$[{(E_8)}_{12}\,\text{SCFT}]\times [{(E_7)}_{24}\times {Spin(7)}_{16}\,\text{SCFT}]$\\
\hline 
17&$\begin{matrix} \includegraphics[width=76pt]{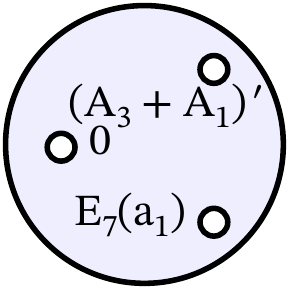}\end{matrix}$&$\{0, 0, 0, 0, 2, 0, 0, 0, 1, 0, 0\}$&$(133, 45)$&$[{(E_8)}_{12}\,\text{SCFT}]\times[{(E_8)}_{24}\times {SU(2)}_{13}\,\text{SCFT}]$\\
\hline 
18&$\begin{matrix} \includegraphics[width=76pt]{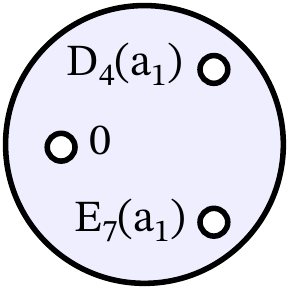}\end{matrix}$&$\{0, 0, 0, 0, 3, 0, 0, 0, 0, 0, 0\}$&$(120, 33)$&${[{(E_8)}_{12}\,\text{SCFT}]}^3$\\
\hline 
19&$\begin{matrix} \includegraphics[width=76pt]{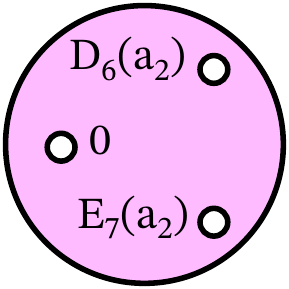}\end{matrix}$&$\{0, 0, 2, 0, 0, 1, 0, 0, 0, 0, 0\}$&$(81, 29)$&$\begin{gathered} [{(E_7)}_{8}\,\text{SCFT}]\\ \times\\ [{(E_7)}_{16}\times {SU(2)}_9\,\text{SCFT}]+\tfrac{1}{2}(56)\end{gathered}$\\
\hline 
20&$\begin{matrix} \includegraphics[width=76pt]{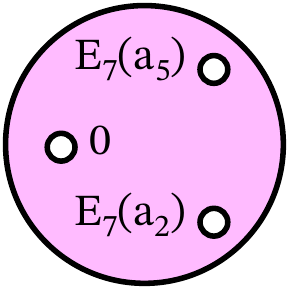}\end{matrix}$&$\{0, 0, 3, 0, 0, 0, 0, 0, 0, 0, 0\}$&$(72, 21)$&${[{(E_7)}_{8}\,\text{SCFT}]}^3+\tfrac{1}{2}(56)$\\
\hline 
21&$\begin{matrix} \includegraphics[width=76pt]{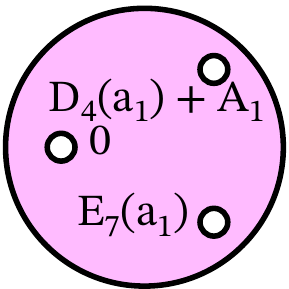}\end{matrix}$&$\{0, 0, 0, 0, 2, 0, 0, 0, 0, 0, 0\}$&$(80, 22)$&${[{(E_8)}_{12}\,\text{SCFT}]}^2+\tfrac{1}{2}(56)$\\
\hline
\end{longtable}
}
\noindent
Here, $n_h$ denotes the effective number of hypermultiplets for interacting part; i.e. it does not include the contribution of the free hypermultiplets in the mixed fixtures \#19--21. These three fixtures contain, in addition, 28 free hypermultiplets transforming as the $\tfrac{1}{2}(56)$ of $E_7$.

The rank-2 ${(E_8)}_{20}$ was given a string theory construction in \cite{Giacomelli:2017ckh}. Various class-S realizations are discussed in section 3.3 of \cite{Chcaltana:2018zag}. The rank-3 ${(E_7)}_{18}\times U(1)$ SCFT was interacting fixture \#13 in \cite{Chacaltana:2015bna}.

Fixture \#19 is obtained from fixture \#20 by replacing the special puncture $E_7(a_5)$ by the $D_6(a_5)$ puncture in the same special piece. The $\mathbb{Z}_2$ Sommers-Achar action has the effect of replacing a pair of rank-1 ${(E_7)}_8$ Minahan-Nemeschansky theories by the rank-2 ${(E_7)}_{16}\times {SU(2)}_{9}$ Minahan-Nemeschansky theory. If, instead, we replaced $E_7(a_5)$ by $A_5+A_1$, the $S_3$ Sommers-Achar action would replace ${[{(E_7)}_8]}^3$ by the rank-3 ${(E_7)}_{24}\times {SU(2)}_{26}$ Minahan-Nemeschansky theory. The same remark holds for the $\mathbb{Z}_2$ Sommer-Achar action relating fixtures \#14 and \#15.

\section*{Acknowledgements}
We would like to thank Mario Martone, Fei Yan and Andrew Neitzke for helpful discussions. This work was supported in part by the National Science Foundation under Grant No. PHY-1620610. 
\bibliographystyle{utphys}
\bibliography{ref}

\end{document}